\documentclass{article}
\usepackage{amsmath,amsfonts,amssymb,amsthm,graphicx}
\usepackage[preprint]{spconf}
\usepackage{url}
\usepackage{hyperref}
\usepackage{caption}
\usepackage{scalefnt}

\usepackage{balance}
\usepackage{bm}

\def\vx{{\bm{x}}}
\def\vy{{\bm{y}}}
\def\vs{{\bm{s}}}

\usepackage{cite}
\usepackage{xcolor}
\usepackage{makecell}
\usepackage{multirow}
\usepackage{etoolbox,siunitx}
\robustify\bfseries
\usepackage{booktabs}

\usepackage{enumitem}
\newcommand\blfootnote[1]{%
  \begingroup
  \renewcommand\thefootnote{}\footnote{#1}%
  \addtocounter{footnote}{-1}%
  \endgroup
}

\makeatletter
\def\blfootnote{\gdef\@thefnmark{}\@footnotetext}
\makeatother

\title{SpecDiff-GAN: A Spectrally-Shaped Noise Diffusion GAN\\for Speech and Music Synthesis}
\name{Teysir Baoueb$^{1}$, Haocheng Liu$^{1}$, Mathieu Fontaine$^{1}$, Jonathan Le Roux$^{2}$, Gaël Richard$^{1}$}
  
\address{$^{1}$LTCI, T\'el\'ecom Paris, IP-Paris, France\\
      $^{2}$Mitsubishi Electric Research Laboratories (MERL), Cambridge, MA, USA}
\copyrightnotice{\copyright 2024 IEEE. Personal use of this material is permitted. Permission from IEEE must be obtained for all other uses, in any current or future media, including reprinting/republishing this material for advertising or promotional purposes, creating new collective works, for resale or redistribution to servers or lists, or reuse of any copyrighted component of this work in other works.}

\begin{document}
\ninept
\maketitle

\begin{abstract}
Generative adversarial network (GAN) models can synthesize high-quality audio signals while ensuring fast sample generation. However, they are difficult to train and are prone to several issues including mode collapse and divergence. In this paper, we introduce SpecDiff-GAN, a neural vocoder based on HiFi-GAN, which was initially devised for speech synthesis from mel spectrogram. In our model, the training stability is enhanced by means of a forward diffusion process which consists in injecting noise from a Gaussian distribution to both real and fake samples before inputting them to the discriminator. We further improve the model by exploiting a spectrally-shaped noise distribution with the aim to make the discriminator’s task more challenging. We then show the merits of our proposed model for speech and music synthesis on several datasets. Our experiments confirm that our model compares favorably in audio quality and efficiency compared to several baselines.
\end{abstract}
\begin{keywords}
Generative adversarial network (GAN), diffusion process, deep audio synthesis, spectral envelope
\end{keywords}
\section{Introduction} \label{sec:intro}
\blfootnote{This work was funded by the European Union (ERC, HI-Audio, 101052978). Views and opinions expressed are however those of the author(s) only and do not necessarily reflect those of the European Union or the European Research Council. Neither the European Union nor the granting authority can be held responsible for them.}
Deep audio synthesis refers to a class of models which leverage neural networks to generate natural-sounding audio signals based on given acoustic features. It has applications in many different tasks including the generation of speech (e.g., text-to-speech (TTS) \cite{shen2018natural}, speech-to-speech translation \cite{Jia2019DirectST}, voice conversion \cite{Sisman2020AnOO}), music synthesis \cite{engel2019gansynth, music_survey}, and sound effects generation \cite{moffat, sound_effect}.

Audio synthesis was for long dominated by likelihood-based models such as autoregressive models \cite{vandenoord16_ssw} and flow-based models \cite{waveglow}. However, the sequential nature of the former models leads to slow inference times as each output element is generated one by one, conditioned on previously generated elements. Flow-based models, on the other hand, are not parameter-efficient as they typically require a deep architecture to perform complex invertible transformations.

With the emergence of generative adversarial networks (GANs) \cite{NIPS2014_gans}, which have yielded promising results in the generation of high-resolution images, GAN-based audio synthesis models have been proposed \cite{kong2020hifigan, lee2023bigvgan}. They can produce high-fidelity waveforms while maintaining a fast and computationally competitive sampling. However, GANs are hard to train and are known to suffer from mode collapse \cite{kodali2018on}. This issue was addressed by denoising diffusion probabilistic models (DDPMs) \cite{NEURIPS2020_ddpm, chen2021wavegrad, lee2022priorgrad}, but these models suffer themselves from a slow reverse process, which requires a huge number of steps to obtain satisfactory results, thus making them inapplicable in real-life settings.

\begin{figure}[tbp]
\centering
\includegraphics[width=.98\columnwidth]{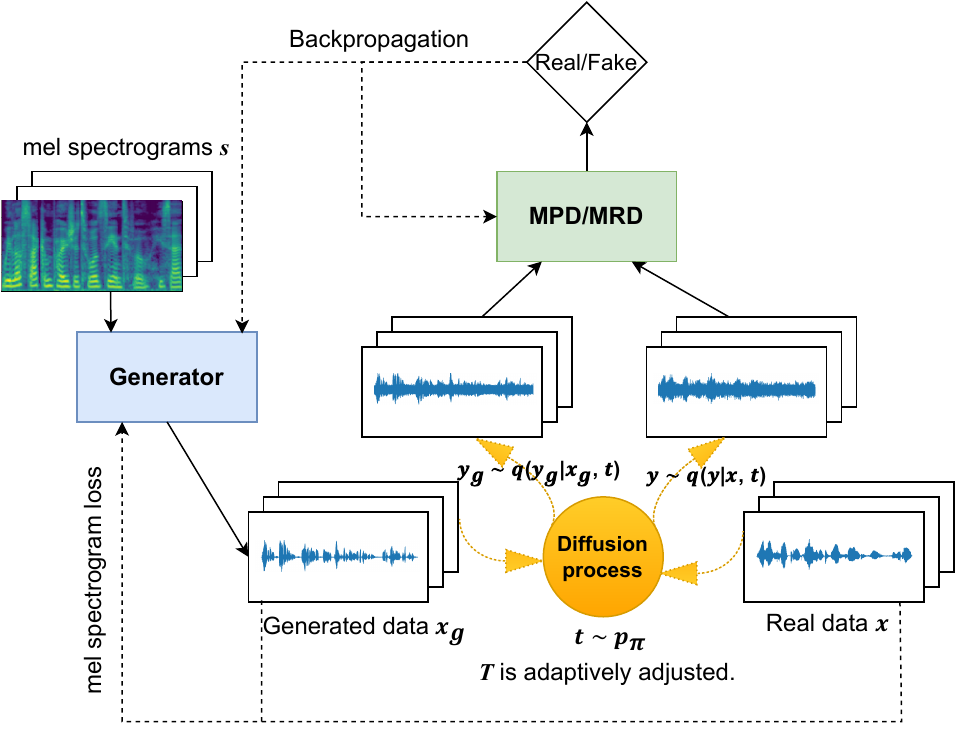}
\caption{Overview of SpecDiff-GAN}
\label{fig:overview}
\vspace{-.3cm}
\end{figure}

In this paper, we propose to tackle the training instability of GANs and the slow inference process of DDPMs. To that aim, we choose HiFi-GAN \cite{kong2020hifigan}, an efficient and high-quality mel spectrogram to speech waveform synthesizer, as a core model, and build an enhanced HiFi-GAN model exploiting a noise-shaping diffusion process, showing the merit of our proposed model on a large variety of audio signals. 
More precisely, our main contributions include:
\begin{itemize}[leftmargin=*]
  \item The injection of instance noise into both inputs (real and fake) of the discriminator similarly to \cite{wang2023diffusiongan} to help stabilize the training;
\item The use of a spectrally-shaped noise distribution to make the discriminator's task more challenging. In particular, we evaluate several variations for the noise distribution exploiting the inverse filter described in \cite{Koizumi2022SpecGradDP}, which is based on the spectral envelope of the mel spectrogram input;
\item An extensive experimental work with application not only to speech but also instrumental music synthesis, which to the best of our knowledge has not been done before for Hifi-GAN based models. 
\end{itemize}
Our proposed model is illustrated in Fig.~\ref{fig:overview}. Examples and full code are available at {\scalefont{0.92} \url{https://specdiff-gan.github.io/}}.

\vspace{-.1cm}
\section{Related work} \label{sec:rel_work}
\vspace{-.1cm}
\subsection{HiFi-GAN}
\vspace{-.1cm}
HiFi-GAN \cite{kong2020hifigan} addresses the challenges of high-quality speech synthesis by leveraging GANs. The model employs a generator network that takes mel spectrograms as input and utilizes a progressive upsampling process to synthesize time-domain waveforms closely resembling the original audio signals. HiFi-GAN's architecture features a multi-receptive field fusion module, which enhances representation by integrating information from different receptive regions. Additionally, it features two discriminators: multi-period discriminator (MPD) and multi-scale discriminator (MSD), which respectively capture periodic patterns and identify long-term dependencies. This approach has demonstrated remarkable performance in generating high-quality audio with improved sampling accuracy and speed.

\subsection{SpecGrad}
SpecGrad, introduced by Koizumi et al.\ \cite{Koizumi2022SpecGradDP}, is a diffusion-based vocoder. This model enhances the quality of synthesized audio by leveraging a diffusion process that adapts the shaping of noise in the spectral domain. Let $\mathcal{N}(0, \bm{\Sigma})$ be the noise distribution. SpecGrad proposes to include information from the spectral envelope into $\bm{\Sigma}$. To achieve this, $\bm{\Sigma}$ is decomposed as $\bm{\Sigma} = \bm{L} \bm{L}^T$, where $\bm{L} = \bm{G}^+ \bm{M}_{\text{SG}}\bm{G}$, with $\bm{G}$ and $\bm{G}^+$ denoting matrix representations of the short-time Fourier transform (STFT) and its inverse, and $\bm{M}_{\text{SG}}$ a complex diagonal matrix representing a filter based on the spectral envelope.  Specifically, the magnitude of $\bm{M}_{\text{SG}}$ aligns with the spectral envelope, while the phase component is obtained as that of the minimum phase response. 
By incorporating spectral envelope information in this way, SpecGrad enhances the modeling of audio signals, resulting in improved audio quality and naturalness in the generated audio compared to previous diffusion models \cite{chen2021wavegrad, lee2022priorgrad}. However, it is important to note that the slow inference speed of SpecGrad limits its suitability for real-world applications.

\subsection{Diffusion-GAN}
Diffusion-GAN \cite{wang2023diffusiongan} is a novel approach for training GANs using diffusion processes to enhance stability and quality. By gradually transforming real and generated samples through an adaptive diffusion process, Diffusion-GAN bridges the gap between initial generator outputs and the target data distribution. This regularization mechanism mitigates challenges associated with mode collapse and unstable training dynamics, contributing to improved training efficiency and sample quality in GANs.
The original paper applied this approach to image synthesis, and its application to the audio domain remains limited
\cite{wu2023enhancing}.

\section{Proposed method} \label{sec:prop_method}
\subsection{Architecture}
Our generator network closely mirrors the architecture used in HiFi-GAN, chosen for its remarkable capability to produce high-quality audio samples swiftly. Furthermore, we incorporate HiFi-GAN's multi-period discriminator (MPD), which comprises several sub-discriminators, each parameterized with a period $p$, to effectively capture periodic patterns. However, instead of utilizing the multi-scale discriminator (MSD), we opted for UnivNet's multi-resolution discriminator (MRD) \cite{Jang2021UnivNetAN}. MRD is a composition of multiple sub-discriminators, each parameterized by a tuple indicating (FFT size, hop size, Hann window length). These varying temporal and spectral resolutions enable the generation of high-resolution signals across the full band. Integrating MRD consistently improves sample quality and reduces artefacts in audio synthesis, as shown in \cite{lee2023bigvgan, wang2022hifiwavegan}.

\subsection{Enhancing the GAN model with diffusion}
\label{sec:diff-gan}
Following \cite{wang2023diffusiongan}, we leverage a diffusion process during GAN training. In this approach, rather than discerning between the original and generated data, the discriminator learns to distinguish between the perturbed versions of each (see Fig. \ref{fig:overview}).

We recall that, during the forward diffusion process, an initial sample denoted as $\vx_0 \sim q(\vx_0)$ undergoes a series of $T$ sequential steps where it is progressively perturbed by Gaussian noise. Denoting the noise schedule by $\{\beta_t\}_{t=1}^T$, this can be formalized as $q(\vx_{1: T} | \vx_{0})=\prod_{t\geq1} q(\vx_{t} | \vx_{t-1})$ with $q(\vx_{t} | \vx_{t-1})=\mathcal{N}(\vx_{t}; \sqrt{1-\beta_{t}} \vx_{t-1}, \beta_{t} \mathbf{I})$. Let $\alpha_t=1-\beta_t$ and $\bar{\alpha}_t = \prod_{u=1}^t \alpha_u$. It can be shown that, in the forward process, $\vx_t$ can be sampled at any arbitrary time step $t$ in closed form by $\vx_t = \sqrt{\bar{\alpha}_t} \vx_0 + \sqrt{1 - \bar{\alpha}_t} \boldsymbol{\epsilon}$, where $\boldsymbol{\epsilon} \sim \mathcal{N}(\mathbf{0}, \bm{\Sigma})$.

In the context of our model, we denote by $\vx \sim p(\vx)$ the ground-truth audio and by $\vs$ the input condition of the generator, i.e., the mel spectrogram of the ground-truth audio. Using these notations, $G(\vs)$ is the generated signal. Perturbed samples are acquired as follows:
\begin{align}
    \vy \sim  q(\vy| \vx, t), \quad &\vy = \sqrt{\bar{\alpha}_t} \vx + \sqrt{1 - \bar{\alpha}_t} \boldsymbol{\epsilon}\\
     \vy_g \sim q(\vy_g| G(\vs), t), \quad &\vy_g = \sqrt{\bar{\alpha}_t} G(\vs) + \sqrt{1 - \bar{\alpha}_t} \boldsymbol{\epsilon}^\prime
\end{align}
where $\boldsymbol{\epsilon}, \boldsymbol{\epsilon}^\prime \sim \mathcal{N}(\bm{0}, \bm{\Sigma})$, $q(\vy| \vx, t)$ is the conditional distribution of the noisy sample $\vy$ given the target data $\vx$ and the diffusion step $t$
and $q(\vy_g| G(\vs), t)$ is the conditional distribution of the noisy sample $\vy_g$ given the generated signal $G(\vs)$ and the diffusion step $t$.

\subsection{Noise distribution}
We explore two options for $\bm{\Sigma}$. In the first case, we set it to $\bm{\Sigma}_{\text{standard}} = \sigma^2 \bold{I}$, a similar approach to that in \cite{wang2023diffusiongan}, where $I$ represents the identity matrix and $\sigma$ is a scalar. We refer to this model as StandardDiff-GAN. In the second option, drawing inspiration from SpecGrad \cite{Koizumi2022SpecGradDP}, we shape the noise based on the spectral envelope. Our filter $\bm{M}_{\text{spec}}$ is however the inverse of the one used in SpecGrad, specifically $\bm{M}_{\text{spec}} = \bm{M}_{\text{SG}}^{-1}$. This choice results in a noise distribution that emphasizes increased noise incorporation in low-energy regions, thereby challenging the discriminator. The version of our model incorporating this noise distribution, with variance $\bm{\Sigma}_{\text{spec}}=\bm{L}_{\text{spec}}\bm{L}_{\text{spec}}^T$ where $\bm{L}_{\text{spec}}= \bm{G}^+ \bm{M}_{\text{spec}}\bm{G}$, is referred to as SpecDiff-GAN.

\subsection{Adaptive diffusion}
Similar to the approach in \cite{wang2023diffusiongan}, we dynamically regulate the level of difficulty for the discriminators during training by incorporating an adaptive update mechanism for the maximum number of diffusion steps, denoted as $T$, within the interval $[T_{\min}, T_{\max}]$. This adaptive adjustment ensures that the discriminators are provided with varying degrees of challenge as they learn to distinguish between real and generated samples. When the discriminators struggle to perform effectively, we decrease $T$ to provide more opportunities for learning from relatively simpler samples, such as non-perturbed or slightly noisy ones. Conversely, if the discriminators find it too easy to differentiate between the diffused generated and real samples, we increase $T$ to introduce more complexity to their task.

To quantify the extent of discriminator overfitting to the training data, we employ a metric similar to that in \cite{karras2020training}, computed over $B$ consecutive minibatches as
\begin{equation}
    r_d = \mathbb{E}[\text{sign}(D_{\text{train}} - 0.5)],
\end{equation}
where $D_{\text{train}}$ represents the discriminator outputs on samples of the training set and $\mathbb{E}[\cdot]$ a mean over the $B$ minibatches. $r_d$ attempts to estimate the portion of the training set for which discriminator outputs would exceed $0.5$. A value of $r_d$ close to $1$ indicates overfitting, while a value close to $0$ suggests no overfitting. We update $T$ every $B=4$ minibatches using the following rule:
\begin{equation}\label{eq:update_t}
    T \leftarrow T + \text{sign}(r_d - d_{\text{target}}) \cdot C,
\end{equation}
where $d_{\text{target}}$ is a hyperparameter representing the desired value for $r_d$, and $C$ is a constant chosen to regulate the rate at which $T$ transitions from $T_{\min}$ to $T_{\max}$.
The diffusion timestep $t \leq T$ is then drawn from a discrete distribution $p_\pi$ defined  with $c_T=\sum_{u=1}^T u$ as:
\begin{align}
    t \sim p_\pi := \text{Discrete}\left(1/c_T, 2/c_T, \dots, T/c_T\right).\label{eq:step}
\end{align}
This distribution gives more weight to larger values of $t$, influencing the choice of diffusion steps during training.

\subsection{Training losses}
We here describe the various training losses. For the sake of simplicity, we denote both discriminators as $D$ following \cite{kong2020hifigan}.

Our discriminative loss is provided by the following formula:
\begin{equation}
    \mathcal{L}_D = \mathbb{E}_{(\vx, \vs, t, \vy, \vy_g)}\left[(D(\vy) - 1)^2 + (D(\vy_g))^2\right],\label{eq:discriminator}
\end{equation}
where $\vy$ and $\vy_g$ are obtained as in Section~\ref{sec:diff-gan}.
For HiFi-GAN, the loss is simply obtained as $\mathcal{L}_D \!=\! \mathbb{E}_{(\vx, \vs)}[(D(\vx) - 1)^2 + (D(G(\vs)))^2]$.

The SpecDiff-GAN generator, as HiFi-GAN, employs an adversarial loss and two extra losses to enhance perceptual and spectral similarity with the ground-truth audio, a feature matching (FM) loss and a mel spectrogram loss. The total loss is formulated as
\begin{align}
    \mathcal{L}_G &= \mathbb{E}_{(\vs, t, \vy_g)}\left[(D(\vy_g) - 1)^2\right]\nonumber\\
    &\phantom{=}+ \lambda_{\text{FM}}\mathbb{E}_{(\vx, \vs, t, \vy, \vs, \vy_g)}\Big[\sum_{i = 1}^L \frac{1}{N_i} \lVert D^i(\vy) - D^i(\vy_g)\rVert_1\Big]\nonumber\\
    &\phantom{=}+  \lambda_{\text{mel}} \mathbb{E}_{(\vx,\vs)} \left[\lVert \phi(\vx) - \phi(G(\vs))\rVert_1\right],\label{eq:generator}
\end{align}
where $\lambda_{\text{FM}}$ and $\lambda_{\text{mel}}$ are scalar coefficients, $\phi$ is a function that transforms a waveform into its mel spectrogram, $L$ denotes the number of layers in the discriminator, $D^i$ the features in the $i^\text{th}$ layer of the discriminator and $N_i$ their number.
It is important to highlight that we employ the diffused versions of the real and fake data only for the first two terms in the generator loss. The last term in Eq.~\eqref{eq:generator} indeed does not involve the discriminator.

\section{Experiments} \label{sec:experiments}
We present hereafter the experimental protocol used to evaluate our method and the baseline models.
\subsection{Datasets}
For our experiments, we consider the following datasets:
\begin{itemize}[noitemsep,topsep=0pt,leftmargin=*]
    \item \textbf{LJSpeech} \cite{ljspeech17} is a 
    single-speaker speech dataset. It contains English recordings sampled at $22050$ Hz with a total duration of $\sim 24$ hours. We use the same train/test split as in HiFi-GAN \cite{kong2020hifigan} (i.e., $12950$ clips for training and $150$ clips for testing).
    \item \textbf{VCTK} \cite{Yamagishi2019CSTRVC} is a clean multispeaker dataset with $110$ speakers, $63$ female and $47$ male. The clips were recorded using two microphones and we consider the Microphone $1$ configuration. It comprises $\sim 41$ hours of utterances in different English accents. We resample the recordings from $48$ kHz to $24$ kHz. We keep $10$ speakers for testing and use the others for training.
    \item \textbf{MAPS} \cite{Emiya2010MAPSA} is a dataset of MIDI piano recordings captured under $9$ distinct recording conditions, and sampled at a rate of $44.1$ kHz. We focus on a specific subset consisting of classical piano compositions (MUS), totaling approximately $18$ hours. We split the dataset into $229$ pieces for training and $41$ pieces for testing. Subsequently, we converted offline all tracks to single-channel audio and segmented them into $5$-second fragments.
    \item \textbf{ENST-Drums} \cite{Gillet2006ENSTDrumsAE} contains recordings by $3$ drummers on $8$ individual audio channels with a total duration of $225$ minutes. The tracks were recorded using various drum kits and are sampled at $44.1$ kHz. We split the recordings into $2512$ for training and $466$ for testing. Similarly to MAPS, our pre-processing pipeline involves an offline conversion from stereo to mono and the subsequent segmentation of audio clips.
\end{itemize}

\subsection{Model Setup}
For MRD, we incorporate $3$ sub-discriminators with the same parameters as \cite{Jang2021UnivNetAN}: $(1024, 120, 600)$, $(2048, 240, 1200)$, and $(512, 50, 240)$. As in \cite{kong2020hifigan}, we consider $5$ sub-discriminators for MPD with periods $2$, $3$, $5$, $7$, and $11$ to prevent overlaps, and $\lambda_{\text{FM}} = 2$ and $\lambda_{\text{Mel}} = 45$ for the generator loss. We also keep the same choice of optimizer and learning rate scheduler. For the diffusion process, we adopt $d_{\text{target}} = 0.6$, where experiments with other values showed no significant difference, and $\sigma = 0.05$ as per \cite{wang2023diffusiongan}.

We compare our model to other GAN models, namely HiFi-GAN\footnote{\url{https://github.com/jik876/hifi-gan}}, UnivNet-c$32$\footnote{\url{https://github.com/maum-ai/univnet}} and BigVGAN base model\footnote{\url{https://github.com/NVIDIA/BigVGAN}}.

\subsection{Training configurations}
Detailed training configurations for each dataset across the various models are as follows:
\begin{itemize}[noitemsep,topsep=0pt,leftmargin=*]
    \item \textbf{LJSpeech:} The parameter values chosen are consistent with the V1 configuration of HiFi-GAN. The initial learning rate is set to $2\cdot 10^{-4}$ across all models, except for UnivNet and BigVGAN, where we conduct experiments with an initial learning rate of $10^{-4}$ in accordance with the settings outlined in their respective papers.
    \item \textbf{VCTK:} We adopt the $24$ kHz base configuration of BigVGAN. The other parameters are the same as those used for LJSpeech.
    \item \textbf{MAPS} and \textbf{ENST-Drums}: 
    We use $128$-dimensional log-mel spectrograms with a Hann window size of $2048$, a frame shift of $512$, and $2048$-point FFT with a full-band range ($0$ - $22.050$ kHz). UnivNet is not used in this configuration due to code adjustments required. For other models, we increase upsampling rates and kernel sizes to $[8,8,2,2,2]$ and $[16,16,4,4,4]$ respectively. All models are trained with an initial learning rate of $2 \cdot 10^{-4}$  and a segment size of $16384$. 
    
\end{itemize}
All models are trained on $1$ NVIDIA A$100$ GPU for $1$M steps, with a batch size of $16$. 
All generators have approximately $14$M parameters.

\section{Results}\label{sec:results}
To evaluate the performance of trained models, we use Perceptual Evaluation of Speech Quality (PESQ) \cite{pesq}, Short-Time Objective Intelligibility (STOI) \cite{stoi} and WARP-Q \cite{Jassim2021WarpQQP} for speech synthesis. For each metric, we report the mean of the scores over all the pieces in the test set. Each 95\% confidence interval around the mean value has margins smaller than $0.03$, $0.001$, and $0.008$ respectively. For music generation, we utilize the Fréchet Audio Distance (FAD) \cite{Kilgour2019FrchetAD} with the VGGish model \cite{vggish} to generate the embeddings.

\subsection{Inference results for the different datasets}
Table \ref{tab:results_LJSeech} presents the results on LJSpeech. SpecDiff-GAN exhibits superior performance in terms of audio quality and speech intelligibility when compared to both the baseline models and BigVGAN. Performance drops with $\bm{\Sigma}_{\text{standard}}$ (StandardDiff-GAN), highlighting the importance of the noise shaping. Our model excels with known speakers in band-limited conditions during inference.

\begin{table}[tbp]
    \centering
    \sisetup{
    detect-weight, 
    mode=text, 
    tight-spacing=true,
    round-mode=places,
    round-precision=3,
    table-format=1.3,
    table-number-alignment=center
    }
 \caption{Inference results on LJSpeech. (lr: initial learning rate)}
 \resizebox{0.92\columnwidth}{!}{
 \begin{tabular}
 {l
    S[round-precision=2,table-format=1.2]@{\,\,}S[round-precision=2,table-format=1.2]
    *{2}{S@{\,\,}S}}
 \toprule
 Model & \multicolumn{1}{c}{PESQ ($\uparrow$)} & \multicolumn{1}{c}{STOI ($\uparrow$)} & \multicolumn{1}{c}{WARP-Q ($\downarrow$)}\\ \midrule
 HiFi-GAN & 3.468 & 0.976 & 1.203\\
 UnivNet (lr=1e-4) & 3.440 & 0.977 & 1.330\\
 StandardDiff-GAN & 3.621 & 0.982& 1.086\\
 SpecDiff-GAN & \bfseries 3.758 & \bfseries 0.985 & \bfseries 1.018\\
 BigVGAN (lr=1e-4) & 3.715 & 0.984 & 1.073\\
\bottomrule
 \end{tabular}
 }
 \vspace{-.2cm}
 \label{tab:results_LJSeech}
\end{table}

\begin{table}[tbp]
    \centering
    \sisetup{
    detect-weight, 
    mode=text, 
    tight-spacing=true,
    round-mode=places,
    round-precision=3,
    table-format=1.3,
    table-number-alignment=center
    }    
 \caption{Inference results on VCTK. (lr: initial learning rate)}
 \resizebox{0.92\columnwidth}{!}{
 \begin{tabular}
 {l
    S[round-precision=2,table-format=1.2]@{\,\,}S[round-precision=2,table-format=1.2]
    *{2}{S@{\,\,}S}}
 \toprule
 Model & \multicolumn{1}{c}{PESQ ($\uparrow$)} & \multicolumn{1}{c}{STOI ($\uparrow$)} & \multicolumn{1}{c}{WARP-Q ($\downarrow$)}\\ \midrule
 HiFi-GAN & 2.965 & 0.937 & 1.213\\
 UnivNet (lr=1e-4) & 3.206 & 0.940 & 1.209\\
 StandardDiff-GAN & 3.368 & 0.955 & 1.046\\
 SpecDiff-GAN & 3.517 & 0.963 & 0.983\\
 BigVGAN (lr=1e-4) & 3.673 & 0.962 & 0.959\\
\bottomrule
 \end{tabular}
 }
 \vspace{-.2cm}
 \label{tab:results_vctk}
\end{table}

\begin{table}[t]
 \centering
\caption{FAD ($\downarrow$) scores on MAPS and ENST-Drums datasets.}
 \begin{tabular}{l c c}
 \toprule
 Model & MAPS & ENST-Drums\\ \midrule
 HiFi-GAN & 0.153 & 0.226\\
 StandardDiff-GAN & 0.108 & \bfseries 0.138\\
 SpecDiff-GAN & 0.080 & 0.149\\
 BigVGAN & \bfseries 0.075 & 0.190\\
\bottomrule
 \end{tabular}
 \vspace{-.2cm}\label{tab:results_maps_drums}
\end{table}

The results for the VCTK dataset, which involves inference on unseen speakers, are reported in Table \ref{tab:results_vctk}. Among the models, BigVGAN with a learning rate of $2 \cdot 10^{-4}$ has the best performance. SpecDiff-GAN closely follows, with a negligible difference that is not statistically significant. It is noteworthy that both SpecDiff-GAN and StandardDiff-GAN outperform the baseline models, HiFi-GAN and UnivNet. In particular, SpecDiff-GAN showcases a substantial performance margin compared to the baselines.

Table \ref{tab:results_maps_drums} displays the results for the MAPS and ENST-Drums datasets. For MAPS, SpecDiff-GAN outperforms both HiFi-GAN and StandardDiff-GAN, highlighting the advantage of employing the spectrally-shaped noise distribution. Notwithstanding, BigVGAN demonstrates a slightly better performance compared to SpecDiff-GAN. Surprisingly, in the case of ENST-Drums, StandardDiff-GAN outperforms the other models, with SpecDiff-GAN following closely behind. 
The ENST-Drums dataset's small size ($225$ minutes) and multiple tracks for the same drum performance from different channels may have hindered the learning process.

\subsection{Ablation study}
We conducted an ablation study on the MRD, the diffusion process, and the reshaped noise distribution to assess the individual impact of each component on the quality of the generated audio. We train all models on LJSpeech for $1$M steps. The results are presented in Table~\ref{tab:results_ablation}. Eliminating the spectrally-shaped noise distribution and adopting $\bm{\Sigma}_{\text{standard}}$ instead (StandardDiff-GAN) leads to a deterioration in results. This behaviour is also observed when replacing the MRD with the MSD. Furthermore, when we exclude the diffusion process entirely (``Without diffusion'')  or maintain it with $\bm{\Sigma}_{\text{standard}}$ and substitute the MSD for the MRD, the results decline even further. It is worth noting that the last row in the table is equivalent to HiFi-GAN with a non-spectrally-shaped diffusion process. A comparison of metric scores with those of HiFi-GAN in Table~\ref{tab:results_LJSeech} reveals that the diffusion process leads to improvement. This highlights that each component of our model (MRD, diffusion, shaped noise) plays a crucial role in enhancing audio quality.

\begin{table}[t]
    \centering
    \sisetup{
    detect-weight, 
    mode=text, 
    tight-spacing=true,
    round-mode=places,
    round-precision=3,
    table-format=1.3,
    table-number-alignment=center
    }
 \caption{Ablation study on LJSpeech.}
 \resizebox{0.92\columnwidth}{!}{
 \begin{tabular}
 {l
    S[round-precision=2,table-format=1.2]@{\,\,}S[round-precision=2,table-format=1.2]
    *{2}{S@{\,\,}S}}
 \toprule
 Model & \multicolumn{1}{c}{PESQ ($\uparrow$)} & \multicolumn{1}{c}{STOI ($\uparrow$)} & \multicolumn{1}{c}{WARP-Q ($\downarrow$)}\\ \midrule
 SpecDiff-GAN & \bfseries 3.758 & \bfseries 0.985 & \bfseries 1.018\\\midrule
 StandardDiff-GAN & 3.621 & 0.982 & 1.086\\
 Without diffusion & 3.524 & 0.979 & 1.135\\
 MRD $\rightarrow$ MSD & 3.645 & 0.982 & 1.069\\
 \makecell{$\left(\bm{\Sigma}_{\text{spec}} \rightarrow \bm{\Sigma}_{\text{standard}}\right)$\\ + (MRD $\rightarrow$ MSD)} & 3.539 & 0.979 & 1.156\\
\bottomrule
 \end{tabular}
 }
 \vspace{-.2cm}\label{tab:results_ablation}
\end{table}

\begin{table}[t]
 \centering
\caption{Synthesis speed compared to real-time evaluated with a batch of $100$ one-second-long samples on $1$ NVIDIA V$100$ GPU}
 \resizebox{\columnwidth}{!}{
 \begin{tabular}{l c c c c}
 \toprule
 Model & LJSpeech & VCTK & MAPS & ENST-Drums\\\midrule
 BigVGAN base & \phantom{1}$\times$ 23.28 & \phantom{1}$\times$ 21.40 & \phantom{1}$\times$ 18.03 & \phantom{1}$\times$ 18.03\\
 SpecDiff-GAN & $\times$ 220.96 & $\times$ 203.28 & $\times$ 183.46 & $\times$ 183.15\\
 \bottomrule
 \end{tabular}}
 \vspace{-.3cm}
\label{tab:inference_speed}
\end{table}

\subsection{Model complexity}
In comparison to the base BigVGAN model, our model features approximately $200$k fewer parameters for all tested configurations. Furthermore, our model demonstrates a notably faster synthesis speed, as detailed in Table \ref{tab:inference_speed}. This speed is equivalent to that of HiFi-GAN and StandardDiff-GAN since they all share the same generator. The primary reason for BigVGAN's slower performance lies in its utilization of the computationally intensive snake activation function \cite{NEURIPS2020snake}. This characteristic also makes BigVGAN significantly slower to train compared to our model, with a training duration factor ranging from $1.5$ to $2$.

\section{Conclusion}\label{sec:conclusion}
We introduced SpecDiff-GAN, a novel approach harnessing a forward diffusion process with spectrally-shaped noise to enhance GAN-based audio synthesis. Our application spanned both speech and music generation. The experimental results showcased SpecDiff-GAN's capacity to generate high-quality waveforms surpassing baselines while being competitive to the state-of-the-art model, BigVGAN. Notably, SpecDiff-GAN maintained efficient inference speeds. Our approach is versatile, offering adaptability to various GAN-based audio synthesis models.

Future research avenues include testing our model on a larger, more diverse dataset, covering a wide spectrum of sound types for universal audio synthesis.

\balance
\vfill\eject
\bibliographystyle{IEEEtran}
\bibliography{refs}

\end{document}